\documentclass{article}
\usepackage{spconf,amsmath,graphicx}
\usepackage{mathtools, cuted} %needed for equation over two columns

\usepackage{tikz}
\usetikzlibrary{external}
\newlength\figureheight 
\newlength\figurewidth 
\usepackage{pgfplots}
\pgfplotsset{compat=newest}

\usepackage{stfloats} %for equation going to the bottom of the page
\usepackage{cite} % to abreviate long enumerations of citated papers
\title{Optimization of a Fixed Virtual Sensing Feedback ANC Controller for In-Ear Headphones with Multiple Loudspeakers}

\name{Piero Rivera Benois$^{1,3}$, Reinhild Roden$^{2}$, Matthias Blau$^{2,3}$ and Simon Doclo$^{1,3}$\thanks{\hspace*{-0.5cm}This research was funded by the Deutsche Forschungsgemeinschaft (DFG, German Research Foundation) – Project-ID 352015383 – SFB 1330 C1.}}
\address{$^{1}$Signal Processing Group, University of Oldenburg, Oldenburg, Germany\\
	$^{2}$Institut für Hörtechnik und Audiologie, Jade Hochschule, Oldenburg, Germany\\
	$^{3}$Cluster of Excellence Hearing4all\\
	{\tt piero.rivera.benois@uni-oldenburg.de} \\
}

\begin{document}
%\ninept
%
\maketitle
\begin{abstract}
In this paper we consider an in-ear headphone equipped with an inner microphone and multiple loudspeakers and we propose an optimization procedure with a convex objective function to derive a fixed multi-loudspeaker ANC controller aiming at minimizing the sound pressure at the ear drum. Based on the virtual microphone arrangement (VMA) technique and measured acoustic paths between the loudspeakers and the ear drum, the FIR filters of the ANC controller are jointly optimized to minimize the power spectral density at the ear drum, subject to design and stability constraints. For an in-ear headphone with two loudspeakers, the proposed multi-loudspeaker VMA controller is compared to two single-loudspeaker VMA controllers. Simulation results with diffuse noise show that the multi-loudspeaker VMA controller effectively improves the attenuation by up to about 10 dB for frequencies below 300 Hz when compared to both single-loudspeaker VMA controllers.
\end{abstract}
\section{INTRODUCTION}
\label{sec:intro}

%ANC in general
Active Noise Cancellation (ANC) applied to headphones aims at minimizing the environmental noise at the listener's ears \cite{simshauser1955}. The working principle of ANC is based on the destructive superposition of sound waves, in this context the sound wave of the environmental noise reaching the listener's ear and the sound wave generated by the loudspeaker of the headphone. Whereas the passive attenuation achieved by the construction materials of the headphone is mainly effective in the mid and high frequencies, ANC is mainly effective in the low frequencies.  %If both sound waves have the same magnitude but opposite phase, the resulting sound pressure is zero. 

%\begin{figure}[]
%	\centering
%	\includegraphics[width=8.25cm]{figures/SISO_FB_VMA_signal_model.png}
%	\caption{Block diagram of the feedback virtual microphone arrangement technique in \cite{pawelczyk2009}. At the left-hand side the DSP of the headphone is shown. At the right-hand side the ear drum is shown}
%	\label{fig:fb_vma_siso}
%\end{figure}

% ANC at the ear drum
Several ANC approaches have been proposed in the literature \cite{kuosbook,elliott2001}, either based on fixed or adaptive feedforward or feedback controllers. In this paper we consider in-ear ANC headphones with an inner microphone and multiple loudspeakers, and we aim at minimizing the sound pressure at the ear drum by means of a fixed multi-loudspeaker feedback ANC controller. Since the sound pressure at the inner microphone and at the ear drum are not equal, several virtual sensing techniques have been proposed (see the literature review in \cite{moreau2008}) to move the zone of quiet from the position of the inner microphone to the position of the ear drum. In \cite{pawelczyk2009} a virtual sensing technique called feedback virtual microphone arrangement (VMA) technique was proposed that exploits the proximity of the inner microphone to the ear drum. Using measured acoustic transfer functions between the loudspeaker and the inner microphone and the ear drum, the VMA technique calculates its control signal using a real-time estimate of the sound pressure at the ear drum with ANC on. 
\begin{figure}[h]
	\centering
	\includegraphics[width=3cm,trim=0 30 0 30,clip]{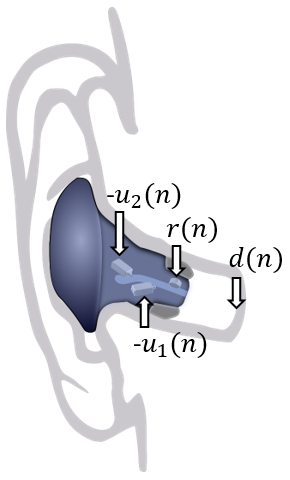}
	\caption{In-ear headphone with an inner microphone and two loudspeakers, $r(n)$ and $d(n)$ denote the signals generated by the incident noise arriving at the inner microphone and the ear drum, respectively, and $-u_1(n)$ and $-u_2(n)$ denote the loudspeaker signals generated by the ANC controller.}
	\label{fig:general_overview}
\end{figure}

In \cite{rafaely1999} and \cite{rivera2020} optimization procedures of an FIR filter as ANC controller have been proposed, which were formulated using objective functions that are convex w.r.t. the FIR filter coefficients. In \cite{rafaely1999} the optimization procedure used the internal model control structure \cite{morari1989} to transform the non-convex objective function of the feedback ANC controller into a convex objective function of an equivalent feedforward ANC controller. In \cite{rivera2020} the optimization procedure approximated the non-convex minimization of the objective function of the feedback ANC controller by the convex maximization of the objective function's denominator. 
In \cite{pawelczyk2009,yu2001,Liang2017,Liang2019,Wang2021} optimization procedures of IIR filters have been proposed, which are formulated using objective functions that are non-convex w.r.t. the IIR filter coefficients. In \cite{pawelczyk2009} the objective function was similarly formulated as in \cite{rivera2020}, but without yielding a convex objective function. The author assumed a twin-biquad filter and applied an exhaustive search to find the optimal parameter values. In \cite{yu2001} a different optimization procedure was proposed that uses the sequential quadratic programming (SQP) method to minimize the non-convex objective function of the feedback ANC controller. In \cite{Liang2017,Liang2019} an open-loop zero-pole placement method was used to design a reasonable initial solution for the optimization. A two-step optimization procedure was proposed in \cite{Wang2021} that uses a genetic algorithm to find an initial solution and the Nelder-Mead simplex method to find the local optimum in its vicinity. All the above mentioned methods make use of only one loudspeaker and, except for \cite{pawelczyk2009}, aim at minimizing the sound pressure at the inner microphone and not at the ear drum. 

% this paper
In this work, we propose an optimization procedure to derive a fixed multi-loudspeaker feedback ANC controller using the VMA technique. To formulate a convex objective function, we assume the ANC controller to be an FIR filter per channel and use the approximation proposed in \cite{rivera2020} to overcome the non-convexity of the objective function. %Additionally, aiming at exploiting the availability of multiple loudspeakers we extend the proposed optimization procedure by assuming the ANC controller to be a single-input multiple-output system with one FIR filter per loudspeaker channel, and jointly optimizing the FIR filters by taking advantage of the convexity of the proposed objective function w.r.t. the filter coefficients of each loudspeaker channel.

%In this work, we use FIR filters to implement the ANC controller and propose a novel optimization procedure by formulating a convex objective function w.r.t. the FIR coefficients using the approach proposed in \cite{rivera2020}. Aiming at minimizing the sound pressure at the ear drum, we extend the optimization procedure in \cite{rivera2020} by using the feedback VMA technique proposed in \cite{pawelczyk2009}. Additionally, aiming at exploiting the availability of multiple loudspeakers we extend the optimization procedure to jointly optimize multiple FIR filters by taking advantage of the convexity of the new objective function w.r.t. all FIR filter coefficients. 

% structure
This paper is structured as follows: Firstly, a description of the virtual sensing technique used in this work will be presented. Secondly, a theoretical analysis of the problem based on the PSD of the signal at the ear drum is made. Thirdly, the proposed optimization procedure for the design of the ANC controller is described. Fourthly, simulations results based on acoustic transfer functions measured with an in-ear ANC headphone prototype are presented to validate the proposed optimization procedure.

\section{FEEDBACK VIRTUAL SENSING}
\label{sec:sig_mod}

% noise sound field
Fig.\,\ref{fig:general_overview} depicts an in-ear headphone equipped with an inner microphone and two loudspeakers. The incident noise arrives first at the inner microphone as $r(n)$ and then at the ear drum as $d(n)$. Based on the inner microphone signal, the multi-loudspeaker feedback ANC controller calculates the signals $-u_1(n)$ and $-u_2(n)$ to drive the loudspeakers, aiming at generating a sound wave that destructively overlaps with the sound wave of the incident noise at the ear drum.
 
%\begin{figure}[h]
%	\centering
%	\includegraphics[width=2.5cm,trim=0 30 0 30,clip]{figures/general_overview_nocolors_SIMO.png}
%	\caption{In-ear headphone, $r(n)$ and $d(n)$ denote the signals generated by the incident noise arriving at the inner microphone and the ear drum, respectively, and $-u_1(n)$ and $-u_2(n)$ denote the loudspeaker signals generated by the ANC controller.}
%	\label{fig:general_overview}
%\end{figure}
%\begin{strip}
\begin{figure*}[bh]
	\begin{equation}
		\label{eq:Phi_ee}
		\Phi_{ee}(f)=\bigg(1 - \frac{|\Phi_{dr}(f)|^2}{\Phi_{dd}(f)\Phi_{rr}(f)}\bigg)\Phi_{dd}(f)
		+ \left|\frac{\Phi_{dr}(f)}{\Phi_{rr}(f)} - \frac{\boldsymbol{W}^\text{T}(f)\boldsymbol{S}(f)}{1 + \boldsymbol{W}^\text{T}(f)\left(\hat{\boldsymbol{S}}(f) + \boldsymbol{B}_r(f) - \hat{\boldsymbol{B}}_r(f)\right)}\right|^2\Phi_{rr}(f),
	\end{equation}
\end{figure*}
%\end{strip}

\begin{figure}[h]
	\centering
	\includegraphics[width=8.25cm]{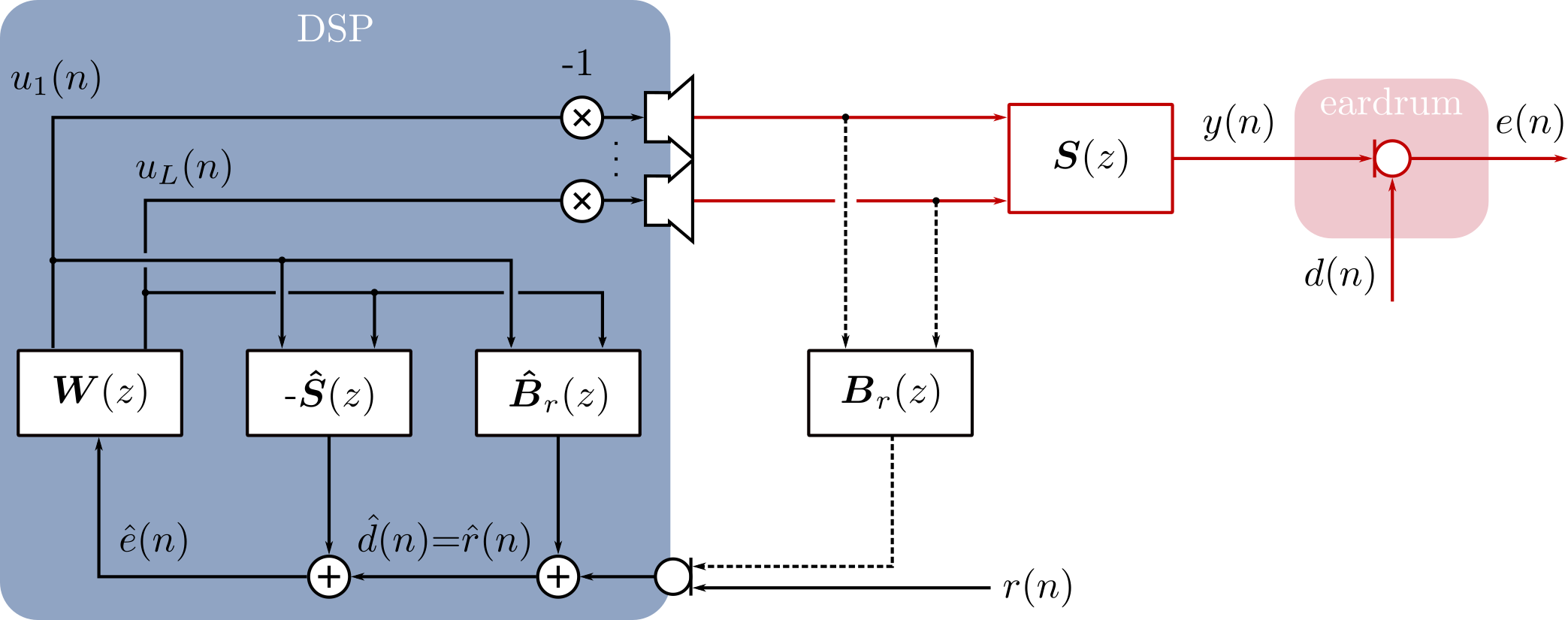}
	\caption{Block diagram of the proposed ANC controller implementing a feedback virtual microphone arrangement technique with multiple loudspeakers.}
	\label{fig:fb_vma}
\end{figure}

The block diagram of the proposed multi-loudspeaker feedback VMA technique is presented in Fig.\,\ref{fig:fb_vma} for the general case of $L$ loudspeakers. Its working principle is the same as the single-loudspeaker VMA technique proposed in \cite{pawelczyk2009}, with the exception that the controller 
\begin{equation}
\boldsymbol{W}(z)=[W_1(z),\ldots,W_L(z)]^\text{T}
\end{equation} 
is a single-input multiple-output system, which generates individual signals for each loudspeaker, and the measured acoustic functions $\boldsymbol{\hat{B}}_r(z)=[\hat{B}_{r,1}(z),\ldots,\hat{B}_{r,L}(z)]^\text{T}$ and $\boldsymbol{\hat{S}}_r(z)=[\hat{S}_1(z),\ldots,\hat{S}_L(z)]^\text{T}$ implement filter-and-sum systems. The technique starts with a calibration stage, in which a probe tube microphone is placed at the ear drum (depicted in pink) to measure the acoustic transfer functions from the loudspeakers to the ear drum $\boldsymbol{S}(z)$ and to the inner microphone $\boldsymbol{B}_r(z)$. Afterwards, the probe tube microphone is removed, the fixed controller $\boldsymbol{W}(z)$ is derived (see Section\,\ref{sec:controller_design}) and the control stage using the DSP (in blue) starts. In the control stage, the sound pressure at the ear drum $e(n)$ is estimated in real-time using the measured acoustic transfer functions $\boldsymbol{\hat{S}}(z)$ and $\boldsymbol{\hat{B}}_r(z)$ in a three-step approach. First, $\boldsymbol{\hat{B}}_r(z)$ is used to estimate $r(n)$ by subtracting the estimated sound pressure generated by the loudspeakers of the headphone at the position of the inner microphone from the microphone signal. Second, the incident noise is assumed to be equal at the inner microphone and at the ear drum, i.e. $\hat{d}(n)=\hat{r}(n)$. Third, the estimated sound pressure at the ear drum $\hat{e}(n)$ with ANC on is estimated by using the estimated sound pressure generated by the incident noise at the ear drum $\hat{d}(n)$ and the measured acoustic transfer functions $\boldsymbol{\hat{S}}(z)$. The resulting signal $\hat{e}(n)$ is then used by the controller $\boldsymbol{W}(z)$ to calculate the control signals $u_1(n)$ to $u_L(n)$.

%Its working principle is the same as the one in \cite{pawelczyk2009}, with the exception that the controllers $\boldsymbol{W}(z)$ generate signals for multiple loudspeakers and that the measured acoustic transfer functions $\boldsymbol{B}_r(z)$ and $\boldsymbol{S}(z)$ are multiple-input single-output systems.
%\begin{figure}[]
%	\centering
%	\includegraphics[width=8.25cm]{figures/SIMO_VMA_block_diagram.png}
%	\caption{Block diagram of the proposed ANC controller implementing a feedback virtual microphone arrangement technique with multiple loudspeakers.}
%	\label{fig:fb_vma}
%\end{figure}

\pagebreak

\section{SYSTEM ANALYSIS}
\label{sec:sys_ana}

%The block diagram of the proposed system is presented in Fig.\,\ref{fig:fb_vma}. The incident noise $r(n)$ is measured by the external microphone. This external microphone signal is used together with the  measured (multi-loudspeaker) acoustic transfer functions $\boldsymbol{B}_r(z)$ and $\boldsymbol{S}(z)$ to calculate the estimated sound pressure at the ear drum while ANC is on $\hat{e}(n)$. This signal is used by the fixed controllers $\boldsymbol{W}(z)$ to calculate the control signals $u_l(n)$ (with l $\in \{1,\ldots, L\}$), which are used to drive the loudspeakers. The generated sound wave travels on the one hand through the acoustic transfer functions $\boldsymbol{B}_r(z)$ to the inner microphone and on the other hand through the acoustic paths $\boldsymbol{S}(z)$ to the ear drum. The signal $e(n)$ at the ear drum is the sum of the control signal $y(n)$ and the incident noise $d(n)$ at the ear drum. 
 %The signal $e(n)$ at the ear drum is the sum of the control signal $y(n)$ generated by the multiple loudspeakers and the incident noise $d(n)$ at the ear drum. 

The power spectral density (PSD) of the sound pressure at the ear drum $e(n)$ at frequency $f$ can be written as in (\ref{eq:Phi_ee}),
%in (\footnote{
%\begin{multline}
%	\label{eq:Phi_ee_ff}
%	\Phi_{ee}(f)=\bigg(1 - \frac{|\Phi_{dr}(f)|^2}{\Phi_{dd}(f)\Phi_{rr}(f)}\bigg)\Phi_{dd}(f)\\
%	\text{+}\left|\frac{\Phi_{dr}(f)}{\Phi_{rr}(f)}\text{ -- }\frac{\boldsymbol{W}^\text{T}(f)\boldsymbol{S}(f)}{1\text{+}\boldsymbol{W}^\text{T}(f)\left(\hat{\boldsymbol{S}}(f)\text{+}\boldsymbol{B}_r(f)\text{--}\hat{\boldsymbol{B}}_r(f)\right)}\right|^2\Phi_{rr}(f),
%\end{multline}
%}),
where $\Phi_{dr}(f)$ denotes the cross-power spectral density (CPSD) between $d(n)$ and $r(n)$, and $\Phi_{dd}(f)$ and $\Phi_{rr}(f)$ denote the PSDs of the respective signals. The left-hand addend determines the minimum achievable PSD $\Phi_{ee}(f)$, which is only equal to zero when the magnitude squared coherence between $d(n)$ and $r(n)$ is equal to 1. The right-hand addend defines the scope of optimization of the controller $\boldsymbol{W}(z)$. Here, the controller $\boldsymbol{W}(z)$ and the acoustic transfer functions $\boldsymbol{B}_r(z)$, $\hat{\boldsymbol{B}}_r(z)$ and $\hat{\boldsymbol{S}}(z)$ build a transfer function, which in series connection with the acoustic paths $\boldsymbol{S}(z)$ should approximate $\Phi_{dr}(z)/\Phi_{rr}(z)$ as well as possible. 

In the VMA technique \cite{elliott1992,pawelczyk2009}, it is assumed that the sound pressure generated by the incident noise is equal at the inner microphone and at the ear drum, i.e. $d(n)=r(n)$. In addition, it is assumed that the measured acoustic transfer functions from the loudspeakers to the ear drum are time-invariant and that after removing the probe tube microphone they remain unchanged, i.e. $\hat{\boldsymbol{S}}(z)=\boldsymbol{S}(z)$. We also assume that changes in the measured acoustic transfer functions to the inner microphone can be compensated with state-of-the-art online estimation methods \cite{Akhtar2013}, i.e. $\hat{\boldsymbol{B}}_r(z)=\boldsymbol{B}_r(z)$. If these assumptions are applied to (\ref{eq:Phi_ee}), the PSD $\Phi_{ee}(f)$ reduces to

\begin{equation}
	\label{eq:Phi_ee_vma}
	\Phi_{ee}^\text{vma} (f)=\bigg|\frac{1}{1+\boldsymbol{W}^\text{T}(f)\hat{\boldsymbol{S}}(f)}\bigg|^2\Phi_{rr}(f).
\end{equation}
Hence, in this case, the scope of optimization of the feedback ANC controller depends only on the PSD of the incident noise at the inner microphone $\Phi_{rr}(f)$ and the measured acoustic transfer functions $\hat{\boldsymbol{S}}(f)$.%, and does not depend on the measured acoustic transfer functions $\hat{\boldsymbol{B}}_r(f)$ as the scope of optimization of the feedback controller in \cite{rivera2020}.

\section{CONTROLLER DESIGN}
\label{sec:controller_design}

%Since the signal at the ear drum can not be directly measured during operation, in this work we use a probe tube microphone placed at the ear drum during a calibration stage to determine the transfer function between the loudspeakers and the ear drum $\boldsymbol{S}(z)$. These acoustic transfer functions are then used for the proposed optimization procedure. 
Designing the controller $\boldsymbol{W}(z)$ by direct minimization of (\ref{eq:Phi_ee_vma}) is a non-convex optimization problem w.r.t. the parameters of $\boldsymbol{W}(z)$. Moreover, a stability constraint is required to restrict the solution space to controllers $\boldsymbol{W}(z)$ that yield a stable system. Similarly as in \cite{rivera2020} for the minimization of the sound pressure at the inner microphone using one loudspeaker, we propose to optimize the multi-loudspeaker controller $\boldsymbol{W}(z)$ by solving the alternative problem of maximizing the denominator in (\ref{eq:Phi_ee_vma}) in the discrete Fourier transform (DFT) domain. %However, we make two changes to the objective function in \cite{pawelczyk2009} to formulate a convex problem 1) we avoid the non-convexity w.r.t. the poles of the IIR filters by assuming $\boldsymbol{W}(z)$ to be a set of FIR filters and 2) we do not calculate the sum over frequency in decibels but instead in squared values. We argue that, when assuming $\boldsymbol{W}(z)$ to be a set of FIR filters, the calculation of the sum over frequency in decibels may yield a solution with single frequencies producing infinite attenuation within a narrow bandwidth. 
The proposed convex objective function is defined as
\begin{equation}
	\label{eq:cost_function}
	\max_\textbf{w} \sum_{k=0}^{L_\text{DFT}/2}\left|1+\boldsymbol{W}^\text{T}(\Omega_k)\hat{\boldsymbol{S}}(\Omega_k)\right|^2G_{1}(\Omega_k),
\end{equation}
where  $L_\text{DFT}$ denotes the DFT length, $\textbf{w}=[\textbf{w}^\text{T}_1,\ldots,\textbf{w}^\text{T}_L]^\text{T}$ denotes the concatenated filter coefficient vector of length ${L\cdot N}$, with $N$ the FIR filter length per channel, $\Omega_k=2\pi k/L_\text{DFT}$ denotes the normalized frequency, $\boldsymbol{W}(\Omega_k)=[W_1(\Omega_k),\ldots,W_L(\Omega_k)]^\text{T}$ contains the frequency response of the ANC controller for each channel, $\hat{\boldsymbol{S}}(\Omega_k)$ contains the frequency response of the measured acoustic transfer functions $\hat{\boldsymbol{S}}(z)$, and $G_1(\Omega_k)$ denotes a frequency weighting function. As in \cite{rivera2020}, we use $G_1(\Omega_k)$ to weight the low frequencies more than the mid and high frequencies.%, and calculate the sum over frequency in squared values and not in dB values, as proposed in \cite{pawelczyk2009}. We argue that, when assuming $\boldsymbol{W}(z)$ to be a set of FIR filters, the calculation of the sum over frequency in decibels may yield a solution with single frequencies producing infinite attenuation. 

Aiming at deriving a controller $\boldsymbol{W}(z)$ that yields a stable system a stability constraint is imposed. Similarly as for the single-loudspeaker case in \cite{rivera2020}, the solution space is restricted to avoid the encirclement of the Nyquist point by means of the following hyperbolic-boundary constraint
\begin{equation}
	\label{ineq:nominal_stability}
	\big|\varrho - \boldsymbol{W}^\text{T}(\Omega_k)\hat{\boldsymbol{S}}(\Omega_k)\big| \leq \big|\varrho +\boldsymbol{W}^\text{T}(\Omega_k)\hat{\boldsymbol{S}}(\Omega_k)|+2\rho  \text{,   } \forall \Omega_k,
\end{equation}
where $\varrho$ determines the focus $(-\varrho,0)$ and $\rho$ determines the x-axis intersect $(-\rho,0)$ of a single-sided hyperbola, with ${0<\rho<1} $ and $\rho<\varrho $. A similar constraint with a fixed focus ${\varrho=1}$ was considered in \cite{polak1984,yu2001}. The advantage of parameterizing the focus $\varrho$ is that after designing the gain margin (GM) using
\begin{equation}
	\text{GM} \geq	\frac{1}{\rho},
\end{equation}
the phase margin (PM) is not fixed and can be designed with $\varrho$ using \cite{rivera2020}
\begin{equation}
	%\text{PM} \geq \arccos\left(\sqrt{\frac{\rho^2\cdot \varrho^2+\rho^2-\rho^4}{\varrho^2}}\right).
\text{PM} \geq \arccos\left(\sqrt{(\rho^2\cdot \varrho^2+\rho^2-\rho^4)/\varrho^2}\right).
\end{equation}
%In \cite{rivera2020} and \cite{yu2001} It has to be stressed that in our case $\boldsymbol{W}(\Omega_k)$ and $\hat{\boldsymbol{S}}(\Omega_k)$ are perfectly known, as they correspond to digital systems inside the DSP. Hence, the GM and PM only have to account for insufficient frequency resolution in the DFT-domain during optimization. 

Feedback ANC approaches are subject to the water-bed effect and therefore prone to produce amplifications outside the attenuation bandwidth, because the associated acoustic transfer functions are typically non-minimum phase systems \cite{pawelczyk2002}. Aiming at restricting the amplification, the design constraint
\begin{equation}
	\label{ineq:nominal_performance}
	 1 \leq G_2(\Omega_k)|1+\boldsymbol{W}^\text{T}(\Omega_k)\hat{\boldsymbol{S}}(\Omega_k)| \text{,   } \forall \Omega_k
\end{equation}
is introduced, where $G_2(\Omega_k)$ denotes the maximum allowed amplification over frequency. In addition, for implementation considerations the gain of the controller is limited by applying the constraint
\begin{equation}
	\label{ineq:max_gain}
	W_l(\Omega_k) \leq G_3(\Omega_k)\text{,   } \forall \Omega_k,
\end{equation}
where the subindex $l  \in\{1,\ldots,L\}$ denotes the loudspeaker channel, and $G_3(\Omega_k)$ denotes the maximum allowed gain over frequency.

The proposed optimization problem maximizing the convex objective function in (\ref{eq:cost_function}), subject to the constraints in (\ref{ineq:nominal_stability}), (\ref{ineq:nominal_performance}) and (\ref{ineq:max_gain}) can then be solved using SQP algorithms, e.g., implemented in the MATLAB function \texttt{fmincon()}.
\pagebreak

\section{SIMULATION RESULTS}
\begin{figure}[h]
	\centering 
%	\includetikz{parameters}	%[width=3cm,trim=0 30 0 30,clip]
	\includegraphics{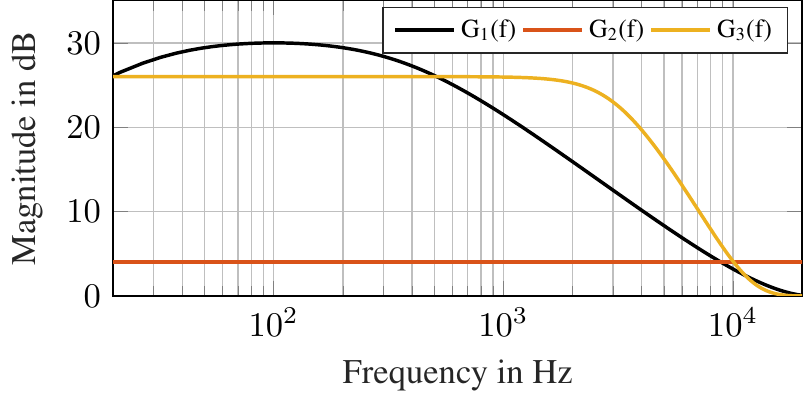}
	\caption{Frequency-dependent functions, where $G_1(f)$ is the frequency weighting in the objective function, $G_2(f)$ is the maximum allowed amplification, and $G_3(f)$ is the maximum controller gain.}
	\label{fig:parameters}
\end{figure}

The performance of the proposed feedback ANC controller was evaluated using the in-ear headphone prototype described in \cite{denk2019, denk2021}, which is equipped with $L=2$ loudspeakers. The acoustic transfer functions between the loudspeakers and the inner microphone $\hat{\boldsymbol{B}_r}(z)$ and the ear drum $\hat{\boldsymbol{S}}(z)$ were measured for a subject wearing the in-ear headphone prototype and using a probe tube microphone. For the simulations, a cylindrically diffuse noise field was considered. All simulations were performed using 10\,s signals at a sampling frequency $f_s=44.1$\,kHz. For the proposed optimization procedure we used $L_\text{DFT}=8192$ in (\ref{eq:cost_function}), and $\rho=0.8$ and $\varrho=2$ in (\ref{ineq:nominal_stability}). Similarly to \cite{rivera2020}, we used frequency-dependent functions $G_1(f)$, $G_2(f)$ and $G_3(f)$ depicted in Fig.\,\ref{fig:parameters}. We compared the performance of the multi-loudspeaker VMA controller, using $N=64$ filter coefficients for each loudspeaker channel, with two single-loudspeaker VMA controllers, using $N=128$ filter coefficients.%and a subject wearing an in-ear headphone equipped with two loudspeakers per ear. Using the acoustic transfer functions measured with an in-ear headphone prototype in \cite{denk2021}, the noise field, the acoustic transfer functions and the DSP algorithms are simulated using a sampling frequency of 44.1\,kHz. For all optimizations the parameters $L_\text{DFT}=8192$, $\rho= 0.8$, $\varrho = 2$ and the frequency-dependent parameters in Fig.\,\ref{fig:parameters} were used. For the proposed multi-loudspeaker VMA technique $L=2$ controllers with $N=64$ filter coefficients each were optimized together and used for a simulation of 10\,s following the block diagram presented in Fig.\,\ref{fig:fb_vma}. Additionally, the single-loudspeaker VMA technique using a controller with $N=128$ filter coefficients was optimized for one channel each time and simulated for 10\,s. As a reference, a fourth simulation is done with ANC off. 
%\begin{figure}[]
%	\centering
%	\includegraphics[width=8.5cm]{figures/Parameters_small.png}
%	\caption{Frequency-dependent optimization parameters, where $G_1(f)$ is the frequency weighting in the objective function, $G_2(f)$ is the maximum allowed amplification, and $G_3(f)$ is the maximum controller gain.}
%	\label{fig:parameters}
%\end{figure}

\begin{figure}[h]
	\setlength{\figureheight}{5.5cm}
	\setlength{\figurewidth}{7cm}
	\centering 
	%\includetikz{psds}
	\includegraphics{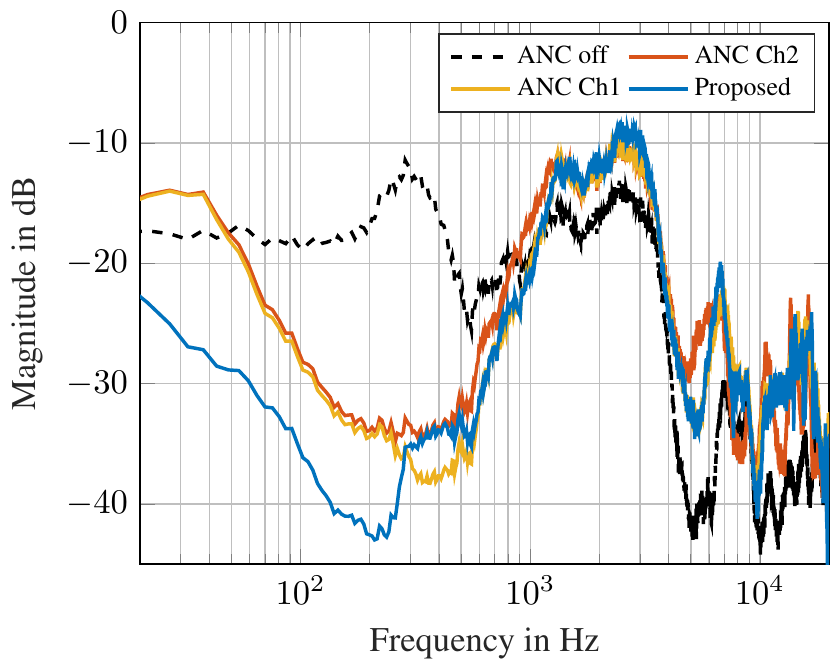}
	\caption{PSD at the left ear of the subject when ANC is off, when the single-loudspeaker VMA technique with either channel 1 or 2 is on, and when the proposed multi-loudspeaker VMA technique (channel 1 and channel 2) is on.}
	\label{fig:results}
\end{figure}

Fig.\,\ref{fig:results} depicts the PSDs of the sound pressure at the ear drum $\Phi_{ee}(f)$ with \texttt{ANC off} and with ANC on, either using single-loudspeaker VMA controllers (\texttt{ANC Ch1}, \texttt{ANC Ch2}) or using the proposed multi-loudspeaker controller (\texttt{Proposed}). It can be observed from \texttt{ANC off} that the passive attenuation of the in-ear headphone is very high for frequencies above 4\,kHz. The single-loudspeaker VMA controllers \texttt{ANC Ch1} and \texttt{ANC Ch2} achieve an attenuation starting at about 45\,Hz and ending at about 800\,Hz or 1\,kHz. The proposed multi-loudspeaker VMA controller clearly outperforms both single-loudspeaker VMA controllers by extending the attenuation bandwidth to frequencies below 20\,Hz and incrementing the attenuation magnitude by about 10\,dB for frequencies below 300\,Hz. For all VMA controllers (single- and multi-loudspeaker) it can be observed that above about 1\,kHz an amplification is produced. Whereas this amplification remains below the maximum amplification of 4\,dB specified by $G_2(f)$ up to about 4\,kHz, it exceeds this value for higher frequencies. This may be due to the fact that the main assumption of the VMA technique, i.e. $d(n)=r(n)$, does not hold in this frequency region, because the wave lengths are comparable to the distance between the inner microphone and ear drum. However, it should be realized that this amplification happens in a frequency region where the passive attenuation of the in-ear headphone is very high. 
%The results are presented in Fig.\,\ref{fig:results}. It can be seen from \texttt{ANC off} that the passive attenuation of the in-ear headphone is very high for frequencies above 4\,kHz. The single-loudspeaker VMA techniques \texttt{ANC Ch1} and \texttt{ANC Ch2} achieve an attenuation that starts at 45\,Hz and ends either at 800\,Hz or at 1\,kHz. Between approx. 1 and 4\,kHz an amplification is produced which remains under the  maximum amplification of 4\,dB specified by $G_2(f)$. Beyond 4\,kHz this amplification is higher than expected, which is the frequency region where the assumption of the VMA technique $d(n)=r(n)$ does not hold anymore, probably because the wave lengths are comparable to the distance between the inner microphone and ear drum. However this happens in a frequency region where the passive attenuation of the in-ear headphone is very high. The proposed multi-loudspeaker VMA technique achieves the best results by extending the attenuation bandwidth to frequencies below 20\,Hz and incrementing the attenuation magnitude by approx. 10\,dB for frequencies below 300\,Hz.
%\begin{figure}[]
%	\centering
%	\includegraphics[width=8cm]{figures/PSDs.png}
%	\caption{PSD at the left ear of the subject when ANC is off, when the single-loudspeaker VMA technique with either channel 1 or 2 is on, and when the proposed multi-loudspeaker VMA technique (channel 1 and channel 2) is on.}
%	\label{fig:results}
%\end{figure}

\section{CONCLUSIONS}
In this paper we considered an in-ear headphone equipped with an inner microphone and multiple loudspeakers and we proposed an optimization procedure to derive a multi-loudspeaker feedback ANC controller aiming at minimizing the sound pressure at the ear drum. We proposed to jointly optimize the FIR filters of the controller by minimizing a convex objective function, subject to design and stability constraints. Using acoustic transfer functions measured with an in-ear headphone prototype with two loudspeakers, the proposed multi-loudspeaker VMA controller was compared to two single-loudspeaker VMA controllers using the same total amount of filter coefficients. Simulation results with diffuse noise field show that the multi-loudspeaker VMA controller effectively improves the attenuation by up to about 10 dB for frequencies below 300 Hz when compared to both single-loudspeaker VMA controllers. In future work the optimization procedure will be extended by integrating uncertainty models of the acoustic transfer functions, e.g., produced by inter-subject variability.

\bibliographystyle{IEEEtr}
\bibliography{refs}

\end{document}